\documentclass[twocolumn,amsmath,amssymb,nofootinbib]{revtex4}

\usepackage{graphicx}

\begin{document}

\title{The role of differential rotation in the evolution of the
\lowercase{$r$}-mode instability}

\author{Paulo M. S\'a}

\email{pmsa@ualg.pt}

\author{Brigitte Tom\'e}

\email{btome@ualg.pt}

\affiliation{Departamento de F\'{\i}sica and Centro
Multidisciplinar de Astrof\'{\i}sica -- CENTRA, \\ F.C.T.,
Universidade do Algarve, Campus de Gambelas, 8005-139 Faro,
Portugal}

\begin{abstract}
We discuss the role of differential rotation in the evolution of
the $l=2$ \textit{r}-mode instability of a newly born, hot,
rapidly-rotating neutron star. It is shown that the amplitude of
the \textit{r}-mode saturates in a natural way at a value that
depends on the amount of differential rotation at the time the
instability becomes active. It is also shown that, independently
of the saturation amplitude of the mode, the star spins down to a
rotation rate that is comparable to the inferred initial rotation
rates of the fastest pulsars associated with supernova
remnants.
\end{abstract}

\maketitle

\section{I\lowercase{ntroduction}}

The \textit{r}-modes are non-radial pulsations modes of rotating
stars that have the Coriolis force as their restoring force and a
characteristic frequency comparable to the rotation speed of the
star \cite{pp}. These modes are driven unstable by gravitational
radiation in all rotating stars \cite{and}. A deeper understanding
of $r$-modes and their astrophysical relevance requires taking
into account nonlinear effects in the evolution of the $r$-mode
instability. One such a nonlinear effect, differential rotation
induced by $r$-modes, has been studied by several authors.
Rezzolla, Lamb and Shapiro \cite{rls} were the first to suggest
that differential rotation drifts of kinematical nature could be
induced by $r$-modes, deriving an approximate analytical
expression for this differential rotation. Afterwards, the
existence of such drifts was confirmed numerically both in general
relativistic \cite{sf} and Newtonian hydrodynamics \cite{ltv}.
Recently, an exact $r$-mode solution, describing differential
rotation of pure kinematic nature that produces large scale drifts
along stellar latitudes, was found within the nonlinear theory up
to second order in the mode's amplitude \cite{sa}. In this paper
we discuss the role of differential rotation in the evolution of
the $l=2$ \textit{r}-mode instability of a newly born, hot,
rapidly-rotating neutron star.

\section{\label{model}T\lowercase{he evolution model of}
O\lowercase{wen \emph{et al}}.}

A few years ago, Owen \emph{et al.} \cite{olcsva} proposed a
simple model to study the evolution of the $r$-mode instability in
newly born, hot, rapidly-rotating neutron stars. Within this
model, it is assumed that the time evolution of the system (star
and $r$-mode perturbation) is characterized by two parameters: the
angular velocity of the star, $\Omega(t)$, and the amplitude of
the mode, $\alpha(t)$. The total angular momentum of the star is
then given by
\begin{equation}
J=I\Omega+J_c,
\end{equation}
where the momentum of inertia $I$ of the equilibrium configuration
is $I =\tilde{I}MR^2$, with $\tilde{I}=0.261$, and the canonical
angular momentum $J_c$ of the \textit{r}-mode perturbation is $J_c
= -3/2 \alpha^2 \Omega \tilde{J} M R^2$, with
$\tilde{J}=1.635\times10^{-2}$. Here, only the $l=2$ $r$-mode is
being considered and it is assumed that the mass density $\rho$
and the pressure $p$ of the fluid are related by a polytropic
equation of state $p=k\rho^2$, with $k$ such that the mass and
radius of the star take the values $M=1.4 M_{\bigodot}$ and
$R=12.53$ km, respectively.

Within this model it is also assumed that the total angular
momentum of the star decreases due to the emission of
gravitational radiation and that the physical energy of the
$r$-mode perturbation (in the co-rotating frame) increases due to
the emission of gravitational radiation and decreases due to the
dissipative effect of viscosity. These assumptions lead then to a
system of differential equations for the angular momentum of the
star, $\Omega$, and the amplitude of the $r$-mode, $\alpha$:
\begin{eqnarray}
\frac{d\Omega}{dt} &=& -\frac{2\Omega}{\tau_{V}}
\frac{Q\alpha^2}{1+Q\alpha^2}, \label{omega}
\\
\frac{d\alpha}{dt} &=& -
\frac{\alpha}{\tau_{GR}}-\frac{\alpha}{\tau_{V}}\frac{1-Q\alpha^2}{1+Q\alpha^2},
\label{alfa}
\end{eqnarray}
where $Q=3\tilde{J}/(2\tilde{I})=0.094$, the
gravitational-radiation and viscous timescales, $\tau_{GR}$ and
$\tau_V$, are given by \cite{lom}
\begin{eqnarray}
\frac{1}{\tau_{GR}} &=& \frac{1}{\tilde{\tau}_{GR}} \left(
\frac{\Omega^2}{\pi G \bar{\rho}} \right)^3, \label{gr-timescale}
\\
 \frac{1}{\tau_V} &=& \frac{1}{\tilde{\tau}_S}
\left( \frac{10^9\mbox{K}}{T} \right)^2 + \frac{1}{\tilde{\tau}_B}
\left( \frac{T}{10^9\mbox{K}} \right)^6 \left( \frac{\Omega^2}{\pi
G \bar{\rho}} \right). \label{v-timescale}
\end{eqnarray}
In the above expressions, the fiducial timescales are
$\tilde{\tau}_{GR}=-3.26\,\mbox{s}$, $\tilde{\tau}_S =
2.52\times10^8\,\mbox{s}$ and
$\tilde{\tau}_B=6.99\times10^8\,\mbox{s}$.

For a newly born, hot, rapidly-rotating neutron star there is an
interval of relevant temperatures and angular velocities of the
star for which the gravitational timescale is much smaller than
the viscous timescale, $\tau_{\rm GR}\ll \tau_{\rm V}$. Therefore,
for this interval of temperatures and angular velocities, we can
neglect in the right-hand side of Eqs.~(\ref{omega}) and
(\ref{alfa}) the terms proportional to $\tau_{V}^{-1}$ and obtain
the solution $\Omega=\Omega_0$ and $\alpha=\alpha_0
\exp\{-(t-t_0)/\tau_{GR}\}$. If the initial angular velocity is
chosen to be $\Omega_0=\Omega_K$, where $\Omega_K=(2/3)\sqrt{\pi
G\bar{\rho}}$ is the Keplerian angular velocity at which the star
starts shedding mass at the equator, then
$\tau_{GR}=-37.1\,\mbox{s}$, implying that the perturbation grows
exponentially from an initial amplitude $\alpha_0=10^{-6}$ to
values of order unity in just about $500\,\mbox{s}$ \cite{olcsva}.
After this first stage of evolution of the $r$-mode instability,
the amplitude $\alpha$ has to be forced, by hand, to take a
certain saturation value $\alpha_{sat}\leqslant Q^{-1/2}=3.26$; in
the second stage of the evolution it is then assumed that
$\alpha=\alpha_{sat}$ and $\Omega$ is determined from the equation
\begin{equation}
\frac{d\Omega}{dt}=\frac{2\Omega}{\tau_{GR}}
\frac{\alpha_{sat}^2Q}{1-\alpha_{sat}^2Q}, \label{first-e1}
\end{equation}
which yields the solution
\begin{eqnarray}
\hspace{-5mm}\Omega(t) &=& \Omega_0 {\biggl [}
\frac{0.030\alpha_{sat}^2}{1- Q \alpha_{sat}^2} \left(
\frac{\Omega_0}{\Omega_K} \right)^6 \left(
\frac{t-t_*}{\mbox{sec}} \right)
 + 1 {\biggr ]}^{-1/6}, \label{omega-k54}
\end{eqnarray}
\vspace{1mm}

\noindent where $t_*$ is the time at which occurs the transition
from the first to the second stage of evolution. During the second
stage of evolution, the star loses its angular momentum, spinning
down to its final angular velocity. As can be seen from the above
equations, the final angular velocity of the star depends
critically on the saturation value of the mode's amplitude
$\alpha_{sat}$; for instance, after one year of evolution, for
$\Omega_0=\Omega_K$ and $\alpha_{sat}=1$ one obtains $\Omega\simeq
0.1\Omega_K$, while for $\alpha_{sat}=10^{-3}$ the angular
velocity is $\Omega\simeq 0.9\Omega_K$. The fact that the growth
of the mode's amplitude has to be stopped by hand at a certain
saturation value introduces an element of arbitrariness into the
solution, permitting, for instance, that agreement between the
predicted final value of the angular velocity of the star and the
value inferred from astronomical observations can always be
achieved by fine-tuning the value of the saturation amplitude. As
will be seen in the next section, this arbitrariness disappears
when differential rotation is taken into account.

\section{\label{model}E\lowercase{volution model with
differential rotation}}

Differential rotation induced by $r$-modes \emph{does} contribute
to the physical angular momentum of the \textit{r}-mode
perturbation \cite{sa}. Therefore, a model for the evolution of
the \textit{r}-mode instability should include the effect of
differential rotation. Here, we adopt the model of Owen \emph{et
al.} \cite{olcsva}, described in the previous section, with the
important difference\footnote{There is another difference, albeit
less significant, between the model of Owen \emph{et al.}
\cite{olcsva} and the model we adopt here, namely, we deduce the
evolution equations just from angular momentum considerations. }
that the total angular momentum of the star is given by
\begin{equation}
J=I\Omega+\delta^{(2)}\!J,
 \label{amtotal}
\end{equation}
where the physical angular momentum of the \textit{r}-mode
perturbation $\delta^{(2)}\!J$ is \cite{sa}
\begin{eqnarray}
\delta^{(2)}\!J &=& \frac12 \alpha^2 \Omega (4K+5) \tilde{J} M
R^2.
 \label{physangm2}
\end{eqnarray}
In the previous expression, $K$ is a constant fixed by the choice
of initial data and gives the initial amount of differential
rotation associated with the \textit{r}-mode. The specific case
$K=-2$, for which the physical angular momentum of the
\textit{r}-mode perturbation coincides with the canonical angular
momentum, corresponds to the model of Owen \emph{et al.}
\cite{olcsva} described in great detail in the previous section.
There is not, to our knowledge, any physical condition that forces
$K$ to take such a particular value $K=-2$. Therefore, we will
consider in this paper arbitrary values of $K$ in the interval
$-5/4\leqslant K\ll 10^{13}$ \footnote{The upper limit for $K$
results from the fact that one wishes to impose the condition that
$|\delta^{(2)}\!J(t_0)|\ll I\Omega_0$ (the initial amplitude
$\alpha_0\equiv\alpha(t_0)$ is considered throughout the paper to
be $10^{-6}$). The lower limit for $K$ results from the fact that
we do not want to saturate the amplitude of the mode by hand, a
procedure needed for the case $K<-5/4$ in order to avoid that the
total angular momentum of the star becomes negative.}.

Following the procedure described in the previous section, we
arrive at a system of two first-order coupled differential
equations determining the time evolution of the amplitude of the
\textit{r}-mode $\alpha(t)$ and of the angular velocity of the
star $\Omega(t)$:
\begin{eqnarray}
\frac{d\Omega}{dt} &=& \frac83 (K+2)Q
\frac{\Omega\alpha^2}{\tau_{GR}}, \label{omega-1}
\\
\frac{d\alpha}{dt} &=& -\left[ 1 + \frac43 (K+2)Q \alpha^2 \right]
\frac{\alpha}{\tau_{GR}}, \label{alfa-1}
\end{eqnarray}
valid when the damping effect of viscosity can be neglected
relatively to the driving effect of gravitational radiation.

\begin{figure}[t]
\centering
\includegraphics[width=0.5\textwidth]{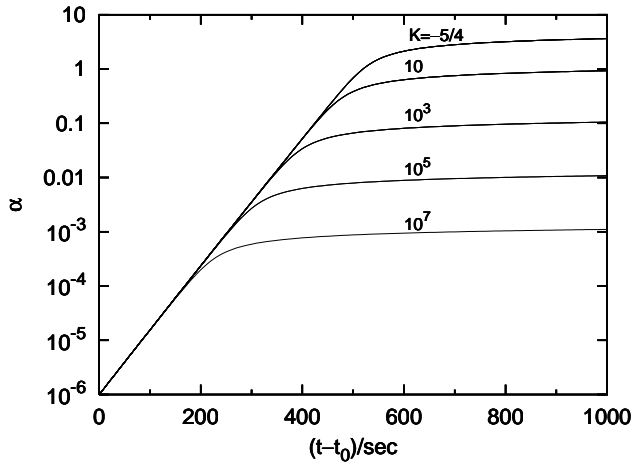}
\caption{Time evolution of the amplitude of the \textit{r}-mode
for different values of $K$. The initial values of the amplitude
of the mode and of the angular velocity of the star are,
respectively, $\alpha_0=10^{-6}$ and $\Omega_0=\Omega_K$.}
\label{fig:alpha}
\end{figure}

This system of equations was solved analytically and discussed in
great detail in Ref.~[9]. In the initial stages of the evolution
of the \textit{r}-mode instability $\alpha$ increases
exponentially \cite{st},
\begin{equation}
\alpha(t) \simeq \alpha_0 \exp \left\{ 0.027 \left(
\frac{\Omega_0}{\Omega_K} \right)^6 \left(
\frac{t-t_0}{\mbox{sec}} \right) \right\}; \label{l-fase}
\end{equation}
while for later times, $\alpha$ increases very slowly as
\begin{equation}
\alpha(t) \simeq 2.48 \left( \frac{\Omega_0}{\Omega_K}
\right)^{3/5} \left( \frac{t-t_0}{\mbox{sec}}
\right)^{1/10}\frac{1}{\sqrt{K+2}}. \label{m-fase}
\end{equation}
The amplitude of the mode saturates in a natural way and a smooth
transition between the regimes (\ref{l-fase}) and (\ref{m-fase})
occurs for $t-t_0\simeq\mbox{few}\times 10^2$ seconds (see
Fig.~\ref{fig:alpha}). This contrasts with the model of Owen
\emph{et al.} \cite{olcsva}, in which, after the short initial
period of exponential growth, the amplitude $\alpha$ has to be
forced by hand to take a certain saturation value
$\alpha_{sat}\leqslant Q^{-1/2}$.
As can be seen from Eq.~(\ref{m-fase}), the saturation value of
the amplitude of the mode depends crucially on the parameter $K$,
namely, $\alpha_{sat}\propto (K+2)^{-1/2}$. If $K\simeq0$,
corresponding to a situation in which the initial amount of
differential rotation is small, then the $r$-mode saturates at
values of order unity. If, on the other hand, the initial
differential rotation associated to \textit{r}-modes is
significant, then the saturation amplitude $\alpha_{sat}$ can be
as small as $10^{-3}-10^{-4}$.

\begin{figure}[t]
\centering
\includegraphics[width=0.5\textwidth]{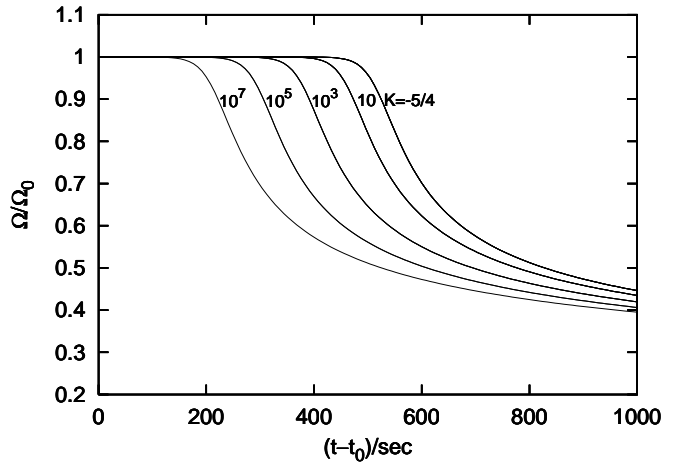}
\caption{Time evolution of the angular velocity of the star for
different values of $K$. The initial values of the amplitude of
the mode and of the angular velocity of the star are,
respectively, $\alpha_0=10^{-6}$ and $\Omega_0=\Omega_K$.}
\label{fig:omega}
\end{figure}

Let us now turn our attention to the time evolution of the angular
velocity of the star, $\Omega$. In the initial stages of the
evolution of the \textit{r}-mode instability $\Omega$ decreases as
\cite{st}
\begin{equation}
\frac{\Omega(t)}{\Omega_0} \simeq 1-\frac43(K+2)Q\alpha_0^2 \exp
\left\{ 0.054 \left( \frac{\Omega_0}{\Omega_K} \right)^6 \left(
\frac{t-t_0}{\mbox{sec}} \right) \right\}; \label{i-fase}
\end{equation}
while for later times $\Omega$ decreases slowly as
\begin{equation}
\frac{\Omega(t)}{\Omega_0} \simeq  1.30 \left(
\frac{\Omega_0}{\Omega_K} \right)^{-6/5} \left(
\frac{t-t_0}{\mbox{sec}} \right)^{-1/5}. \label{f-fase}
\end{equation}
The smooth transition between the regimes (\ref{i-fase}) and
(\ref{f-fase}) occurs for $t-t_0\simeq\mbox{few}\times 10^2$
seconds (see Fig.~\ref{fig:omega}).
Remarkably, in the later phase of the evolution, the angular
velocity $\Omega$ does not depend on the value of $K$ and,
consequently, does not depend on the saturation value of $\alpha$.
This contrasts with the results obtained in Ref.~[7], where the
value of $\Omega$ depends critically on the choice of
$\alpha_{sat}$.

After about one year of evolution, when the dissipative effect of
viscosity becomes dominant and starts damping the mode, the
angular velocity of the star becomes $\Omega_{\mbox{\small{one
year}}}\simeq 0.042 \Omega_K$ (for $\Omega_0=\Omega_K$), in good
agreement with the inferred initial angular velocity of the
fastest pulsars associated with supernova remnants.

\section{\label{con}C\lowercase{onclusions}}

In this paper we have discussed the role of differential rotation
in the evolution of the $l=2$ \textit{r}-mode instability of a
newly born, hot, rapidly-rotating neutron star. We have shown
that, a few hundred seconds after the mode instability sets in,
the amplitude of the \textit{r}-mode saturates in a natural way at
values that depend on the initial amount of differential rotation
associated to the $r$-mode. If the initial differential rotation
of \textit{r}-modes is small, then the \textit{r}-mode saturates
at values of order unity. On the other hand, if the initial
differential rotation is significant, then the saturation
amplitude can be as small as $10^{-3}-10^{-4}$. These low values
for the saturation amplitude of \textit{r}-modes are of the same
order of magnitude as the ones obtained in recent investigations
on wind-up of magnetic fields \cite{rls} and on nonlinear
mode-mode interaction \cite{afmstw}. We have also shown that the
value of the angular velocity of the star becomes, after a short
period of evolution of the \textit{r}-mode instability, very
insensitive to the saturation value of the mode's amplitude. After
about one year of evolution the angular velocity is
$0.042\Omega_K$ (for any $\alpha_{sat}$), in good agreement with
the inferred initial angular velocity of the fastest pulsars
associated with supernova remnants.

\section*{Acknowledgments}
We thank \'Oscar Dias and Luciano Rezzolla for helpful
discussions. This work was supported in part by the \emph{Funda\c
c\~ao para a Ci\^encia e a Tecnologia} (FCT), Portugal. BT
acknowledges financial support from FCT through grant PRAXIS
XXI/BD/21256/99.


\begin{thebibliography}{00}
\bibitem{pp} J. Papaloizou and J. E. Pringle,
{\it Mon. Not. R. Astron. Soc.} {\bf 182}, 423 (1978).

\bibitem{and} N. Andersson,
{\it Astrophys. J.} {\bf 502}, 708 (1998).

\bibitem{rls} L. Rezzolla, F. K. Lamb and S. L. Shapiro,
{\it Astrophys. J.} {\bf 531}, L139 (2000).

\bibitem{sf} N. Stergioulas and J. A. Font,
{\it Phys. Rev. Lett.} {\bf 86}, 1148 (2001).

\bibitem{ltv} L. Lindblom, J. E. Tohline and M. Vallisneri,
{\it Phys. Rev. Lett.} {\bf 86}, 1152 (2001).

\bibitem{sa} P. M. S\'a, {\it Phys. Rev. D.} {\bf 69}, 084001 (2004).

\bibitem{olcsva} B. J. Owen, L. Lindblom, C. Cutler,
B. F. Schutz, A. Vecchio and N. Andersson, {\it Phys. Rev. D.}
{\bf 58}, 084020 (1998).

\bibitem{lom} L. Lindblom, B. J. Owen and S. M. Morsink,
{\it Phys. Rev. Lett.} {\bf 80}, 4843 (1998).

\bibitem{st} P. M. S\'a and B. Tom\'e,
{\it Phys. Rev. D} {\bf 71}, 044007 (2005).

\bibitem{afmstw} P. Arras, E. E. Flanagan, S. M. Morsink,
A. K. Schenk, S. A. Teukolsky and I. Wasserman, {\it Astrophys.
J.} {\bf 591}, 1129 (2003).

\end{thebibliography}
\end{document}